\documentclass[a4paper,12pt]{article}
\usepackage{complexity}
\usepackage{lineno}
\usepackage{fullpage}
\usepackage{hyperref}
\usepackage{setspace}
\usepackage{tikz}
\usepackage{xcolor}
\usepackage{enumerate}
\usepackage[T1]{fontenc}
\usepackage{lmodern}
\usepackage{graphicx}
\usepackage{textgreek}
\usepackage{indentfirst}
\usepackage{pstricks,pspicture}
\usepackage{float}
\usepackage{amsthm,amsmath,amssymb}
\usepackage{pst-plot}
\usepackage{epsfig}
\usepackage{setspace}
\usepackage{float}
\usepackage{rotating}
\usepackage[left=2cm,top=2cm, bottom=2cm,right=2cm]{geometry}

\title{Domination in Graph Theory: A Bibliometric Analysis of Research Trends, Collaboration and Citation Networks}\date{}

\author{Jonecis~A.~Dayap$^1$, Leomarich~F.~Casinillo$^2$, Bijo~S.~Anand$^3$, \\Joey~S.~Estorosos$^1$ and Ricky~B.~Villeta$^4$
       }

\begin{document}
\maketitle
\begin{center}
    $1$ Department of Mathematics and Sciences, University San Jose-Recoletos, Cebu, Philippines \\
           jdayap@usjr.edu.ph, joey.estorosos@usjr.edu.ph\\
    $2$ Department of Mathematics, Visayas State University, Leyte, Philippines\\
leomarichcasinillo02011990@gmail.com\\
$3$ Department of Mathematics, Sree Narayana College, Punalur, Kerala, India\\
bijos\_anand@yahoo.com
\\$4$ Department of Mathematics, Cebu Normal University, Cebu, Philippines\\
rbvilleta35@usjr.edu.ph
          
\end{center}

\begin{abstract}This study conducts a comprehensive bibliometric analysis of research on domination in graph theory from 1961 to 2024, based on Scopus-indexed publications retrieved using the query (“dominating” OR “domination”) AND “graph.” The analysis examines publication trends, key contributors, collaboration patterns, citation impact, and emerging research themes. Results indicate a significant and sustained increase in research output, particularly in recent decades. Henning, M.A., Hedetniemi, S.T., and Haynes, T.W. are identified as the most highly cited researchers, underscoring their foundational contributions to the field. Co-authorship network analysis reveals strong international collaborations, with Sheikhholeslami, S.M. exhibiting the highest total link strength, while the United States emerges as the leading hub for global research partnerships. Keyword co-occurrence analysis identifies four major research clusters: graph algorithms, graph-theoretic foundations, domination variants, and binary graph operations. Notably, recent studies increasingly focus on how domination properties evolve under different graph operations. Citation network analysis confirms the enduring influence of foundational studies while highlighting a shift towards computational and applied methodologies. These findings highlight the transition from theoretical to applied research, emphasizing the role of advanced algorithms, interdisciplinary approaches, and large-scale computational techniques. Future research directions should explore machine learning-based optimization, domination in evolving networks, and applications in cybersecurity, bioinformatics, and large-scale social networks.



\end{abstract}

\section{Introduction }
Graph theory has been a fundamental area of study in discrete mathematics, with domination in graphs emerging as a critical subfield due to its wide range of applications in network security, social networks, bioinformatics, and optimization problems \cite{1,2}.  A dominating set in a graph is a subset of vertices such that every vertex in the graph is either in this set or adjacent to at least one vertex in the set, making it a crucial concept for efficient network control and resource allocation \cite{3}. Given its significance, research on domination in graph theory has expanded substantially, particularly in algorithmic development, complexity analysis, and real-world applications. This study aims to conduct a bibliometric analysis of domination in graph theory, examining research trends, key collaborations, and citation networks from 1961 to 2024 based on data extracted from the Scopus database. By mapping the evolution of this research field, the study provides insights into influential authors, high-impact publications, and thematic trends that have shaped its development.
\par
The formal study of domination in graphs began in the late 1950s when Claude Berge introduced the concept of the coefficient of external stability, which laid the foundation for domination theory. Later, Oystein Ore in 1962 formalized a concept now recognized as the domination number of a graph. A significant advancement in the field occurred in 1977 when Cockayne and Hedetniemi published a comprehensive analysis of dominating sets, consolidating existing findings and expanding the theoretical framework \cite{4}. Over the decades, research on domination in graphs has evolved, leading to the development of numerous variations and extensions tailored to address specific challenges in network optimization and computational problems. Variants such as total domination, independent domination, and secure domination have been explored for their applicability in modeling real-world systems \cite{5}. In parallel, bibliometric analyses have been conducted to explore collaboration structures and knowledge diffusion within the broader domain of graph theory \cite{6,7}. For instance, the study in \cite{6} provides a comprehensive analysis of research trends, research capabilities, and emerging hotspots in graph theory. Similarly, the bibliometric analysis in \cite{7} explores the interdisciplinary applications of graph theory.
\vspace{5pt}
\par
Despite the growing body of literature on domination in graphs, there is still a lack of systematic evaluation of its research impact and thematic evolution over time. Most existing studies focus on theoretical advancements and algorithmic contributions without assessing the research dynamics through bibliometric techniques. Additionally, while some bibliometric analyses exist for general graph theory \cite{6, 7}, they do not provide an in-depth exploration of domination-related research in terms of citation trends, collaborative networks, and keyword evolution. The absence of such an analysis makes it difficult to identify the most influential contributors, institutions, and research themes driving this field forward.
\vspace{5pt}
\par
To address this gap, this study employs bibliometric methods to analyze research trends in domination in graph theory from 1961 to 2024. Using data retrieved from the Scopus database with the keywords "(dominating OR domination) AND graph," this paper applies co-authorship, co-citation, and keyword co-occurrence analyses to identify influential researchers, highly cited works, and thematic progressions. By visualizing collaboration networks and research clusters, this study aims to provide a structured overview of the field's development and potential future research directions. The findings may offer valuable insights for researchers seeking to explore new frontiers in domination theory and its interdisciplinary applications. Moreover, the results of this study may be used as baseline information for future researchers and are hoped to contribute to global impact in the literature of domination in graphs.

\section{Methods}
\subsection{Data Collection}
The data for this bibliometric analysis were retrieved from the Scopus database, a widely recognized indexing platform for scientific literature. The search was conducted within the title, abstract, and keywords fields using the Boolean query (domination" OR "dominating") AND "graph" to ensure comprehensive coverage of relevant studies. The search covered the period from 1961 to 2024 and was restricted to journal articles published in English. A total of 7,324 journal articles were collected. The dataset was extracted on February 8, 2025, and the records were exported in the CSV file format. Each document record included the document title, author, keywords, year of publication, source title, affiliation, citation count, and other relevant bibliographic information. 
\subsection{Data Analysis}
This study employs bibliometric analysis to examine the research landscape of domination in graph theory, with a focus on identifying key trends and scholarly contributions in the field. The primary objective is to analyze research output trends over time, key contributors, and global research distribution. Specifically, the study investigates publication and citation trends, leading authors, country-level research contributions, co-authorship networks, and keyword co-occurrence patterns to map the intellectual structure of domination in graph theory. To facilitate data processing and visualization, VOSviewer 1.6.20 and the Scopus analytical tool were utilized for network analysis and trend identification.

\section{Results}
\subsection{Research Output Trends over Time}
\begin{figure}
    \centering
\includegraphics[width=0.5\linewidth]{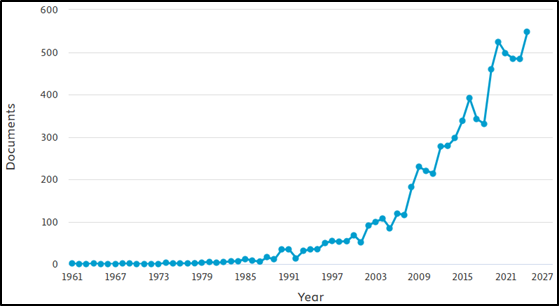}
    \caption{Number of annual research publications from 1961 – 2024}
    \label{fig:1}
\end{figure}

Figure \ref{fig:1} illustrates the annual publication trend of research on domination in graph theory from 1961 to 2024, revealing a steady increase in research output, particularly from the early 2000s onward. The number of publications remained low from 1961 to the late 1980s, followed by a gradual increase in the 1990s. A significant rise in publication output was observed in the early 2000s, with an accelerated growth rate after 2010. 
\subsection{Key Contributors in Domination in Graph Theory}
\subsubsection{Leading Researchers}
\begin{figure}
    \centering
\includegraphics[width=0.55\linewidth]{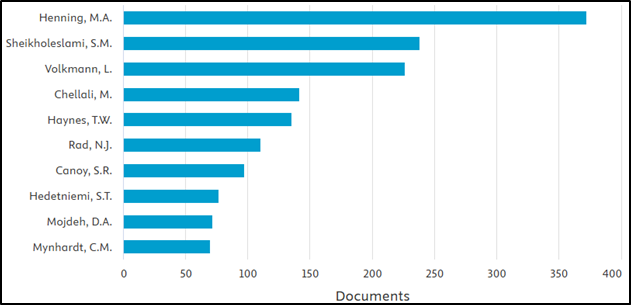}
    \caption{Top 10 Researchers in Domination in Graph Theory Based on Publication Output}
    \label{fig:2}
\end{figure}

Figure \ref{fig:2} presents the top 10 researchers in domination in graph theory, ranked by the number of Scopus-indexed publications. Henning, M.A. leads the field with nearly 400 publications, followed by Sheikhholeslami, S.M. and Volkmann, L., each with over 200 published documents. Other notable contributors include Chellali, M., Haynes, T.W., Rad, N.J., Canoy, S.R., Hedetniemi, S.T., Mojdeh, D.A., and Mynhardt, C.M., all of whom have published more than 50 research papers, reflecting their significant contributions to the field.
\subsubsection{Leading Countries}
\begin{figure}
    \centering
    \includegraphics[width=0.5\linewidth]{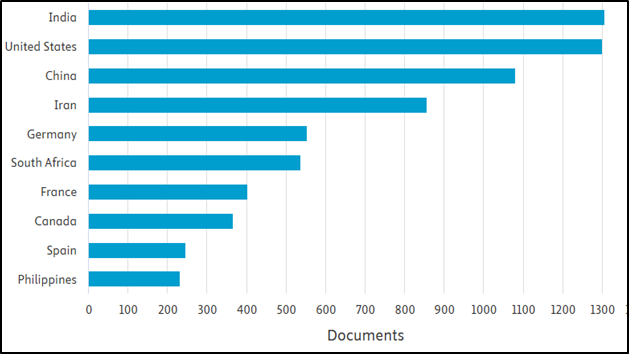}
    \caption{Top 10 Countries in Domination in Graph Theory Based on Publication Output}
    \label{fig:3}
\end{figure}

Figure \ref{fig:3} illustrates the leading countries in domination in graph theory research, ranked by the number of Scopus-indexed publications. India emerges as the top contributor, with over 1,300 publications, followed closely by the United States. Other key contributors include China, Iran, Germany, South Africa, France, Canada, Spain, and the Philippines, reflecting the global research engagement in this domain.
\subsubsection{Leading Source}
\begin{figure}
    \centering
    \includegraphics[width=0.5\linewidth]{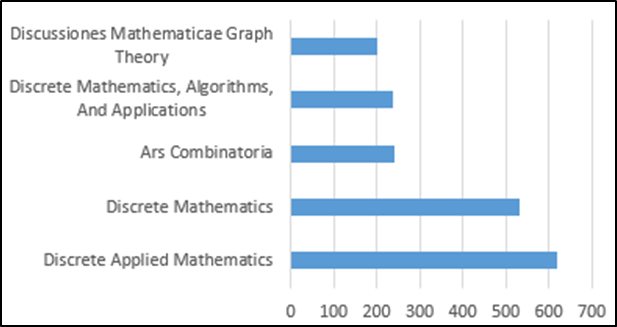}
    \caption{Top 5 Journals in Domination in Graph Theory Based on Publication Output}
    \label{fig:4}
\end{figure}

Figure \ref{fig:4} presents the top five sources for research on domination in graph theory, based on the number of Scopus-indexed publications. Discrete Applied Mathematics leads with 618 publications, followed by Discrete Mathematics with 531 publications. Ars Combinatoria ranks third with 242 publications, while Discrete Mathematics, Algorithms, and Applications and Discussiones Mathematicae Graph Theory contribute 237 and 200 publications, respectively. These sources represent key publication venues for advancements in domination theory and its applications in graph theory.
\subsection{Co-Authorship Network Analysis}
\begin{figure}
    \centering
\includegraphics[width=1.0\linewidth]{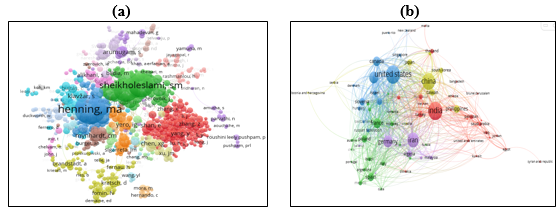}
    \caption{Co-Authorship Network Analysis by (a) Authors and (b) Countries}
    \label{fig:5}
\end{figure}

Figure \ref{fig:5}(a) presents the co-authorship network analysis, highlighting key researchers based on collaborative strength in domination in graph theory. Sheikhholeslami, S.M. has the highest total link strength (562), indicating the most extensive research collaborations, followed by Henning, M.A. (508) and Chellali, M. (296). These results suggest that Sheikhholeslami, S.M. is the most interconnected researcher in the field, demonstrating a broad and active research network, while Henning, M.A. and Chellali, M. also maintain strong collaborative ties within the academic community. Figure \ref{fig:5}(b) presents the co-authorship network at the country level, identifying key nations based on collaborative strength in domination in graph theory. The analysis includes 138 countries, with the United States exhibiting the highest total link strength (894), followed by Iran (558), South Africa (503), China (492), and Germany (456). These findings indicate that the United States has the most extensive international research collaborations in the field, while Iran, South Africa, China, and Germany also demonstrate significant academic networks and cross-country partnerships.  
\subsection{Keyword Co-Occurrence Analysis}
\begin{figure}
    \centering \includegraphics[width=1.0\linewidth]{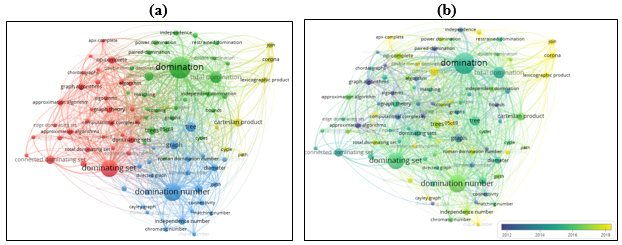}
    \caption{Keyword Co-Occurrence Network in Domination in Graph Theory: (a) Network Visualization and (b) Overlay Visualization}
    \label{fig:6}
\end{figure}

Figure \ref{fig:6}(a) presents the keyword co-occurrence network, identifying major research themes in domination in graph theory. The network consists of four main research clusters, with node sizes representing keyword frequency and edges indicating relationships between terms. The first cluster (red) focuses on algorithmic and computational aspects of domination, with prominent keywords including "dominating set," "connected dominating set," "graph algorithms," and "approximation algorithm." The second cluster (blue) highlights graph-theoretic foundations and domination parameters, featuring terms such as "domination number," "minimum dominating set," "chromatic number," and "diameter." The third cluster (green) explores variants and applications of domination in different graph classes, with key terms like "total domination," "restrained domination," "secure domination," and "Roman domination." The fourth cluster (yellow) focuses on binary operations in graphs, including "Cartesian product," "lexicographic product," "corona," and "join." These clusters collectively represent the dominant research directions in domination theory, emphasizing computational methods, theoretical properties, graph variants, and structural operations. Figure \ref{fig:6}(b) presents the overlay visualization, highlighting recent research trends in domination in graph theory. The yellow-colored nodes, which correspond to binary operations in graphs (e.g., Cartesian product, lexicographic product, corona, and join), indicate a growing focus on their role in domination-related problems. This suggests an increasing research interest in how domination properties evolve under different graph operations.
\subsection{Citation Network Analysis}
\begin{figure}
    \centering
    \includegraphics[width=1.0\linewidth]{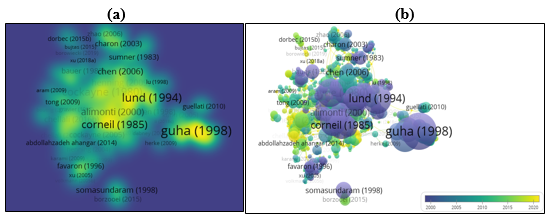}
    \caption{Citation Network Analysis of Research Documents on Domination in Graph Theory: (a) Density Visualization and (b) Temporal Overlay Visualization}
    \label{fig:7}
\end{figure}

Figure \ref{fig:7}(a) presents the density visualization of the citation network analysis, where high-density regions (bright yellow areas) indicate documents with a high citation frequency, signifying their foundational role in domination in graph theory. The most influential works include Guha (1998), "Approximation Algorithms for Connected Dominating Sets," which has 940 citations, highlighting its significant contribution to algorithmic approaches in domination problems; Lund (1994), "On the Hardness of Approximating Minimization Problems," with 613 citations, emphasizing its impact on complexity theory; and Stojmenovic (2002), "Dominating Sets and Neighbor Elimination-Based Broadcasting Algorithms in Wireless Networks," with 605 citations, demonstrating its relevance in network communication and optimization. Additionally, Cockayne (1980), "Total Domination in Graphs," with 534 citations, remains a fundamental study in total domination theory.
\\
\vspace{5pt} 
Figure \ref{fig:7}(b) provides the overlay visualization, illustrating the temporal evolution of research contributions, where the color gradient represents the average publication year—blue nodes correspond to older foundational studies, while yellow-green nodes highlight emerging studies gaining recent attention. The emerging research trends, as indicated by these yellow-green nodes, suggest increasing focus on advanced algorithmic approaches, including machine learning-driven optimization and metaheuristic techniques for NP-hard domination problems. Furthermore, studies on domination in complex and dynamic networks, such as temporal graphs, multiplex networks, and heterogeneous structures, are becoming more prominent. The application of domination concepts in wireless sensor networks, cybersecurity, and social network analysis is also expanding, particularly in network resilience, misinformation detection, and influence spread modeling. Additionally, recent studies explore new domination variants, such as secure domination and defensive domination, along with investigations into binary operations on graphs, including Cartesian product, lexicographic product, and corona operations.
\subsection{Co-citation network analysis}
\begin{figure}
    \centering
    \includegraphics[width=1.0\linewidth]{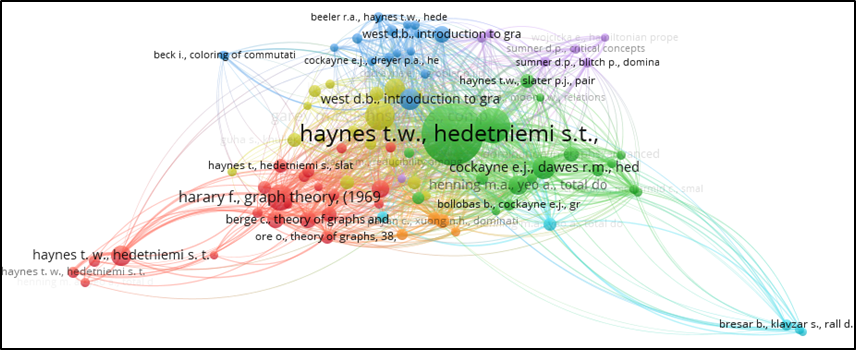}
    \caption{Co-Citation Network Analysis of Cited References in Domination in Graph Theory}
    \label{fig:8}
\end{figure}

The co-citation network analysis as presented in Figure \ref{fig:8} identifies key foundational works in domination in graph theory, forming a structured network of 7 distinct clusters from 88 cited references that met the minimum threshold of 30 citations out of 94,983 total cited references. The most central reference is Fundamentals of Domination in Graphs (TLS = 2,291) by Haynes, Hedetniemi, and Slater (1998). Domination in Graphs: Advanced Topics (TLS = 1,132) by the same authors exhibits strong co-citation links. Total Domination in Graphs (TLS = 624) by Cockayne, Dawes, and Hedetniemi (1980) demonstrates frequent co-occurrence with studies on domination variants. Computers and Intractability: A Guide to the Theory of NP-Completeness (TLS = 525) by Garey and Johnson (1979) presents strong linkages with computational complexity.
\section{Discussion}
The steady growth in research output on domination in graph theory, particularly after the early 2000s, suggests a rising interest in both theoretical and applied aspects of domination-related problems. This surge aligns with the increasing use of graph-theoretic approaches in network security, wireless communication, and combinatorial optimization \cite{8, 9, 10}. The exponential rise in publications after 2010 indicates that domination in graphs has evolved beyond pure mathematics, finding relevance in interdisciplinary domains \cite{11, 12, 13, 38, 39, 40}. The recent fluctuations in publication rates may reflect shifts in research focus, with newer studies exploring variants of domination \cite{32, 33, 34, 35} and emerging applications in dynamic and temporal networks \cite{14, 15}. Moreover, there are also graph theorists who contributed to the foundational dominating sets which include a variety of in-depth mathematical proofs and elegant techniques used in domination theory \cite{31}.
\vspace{5pt}
\par
The dominance of Henning, M.A., Sheikhholeslami, S.M., and Volkmann, L. in terms of research productivity suggests that theoretical advancements in domination remain a highly active area \cite{4}. The high citation counts of Henning, M.A. (6,123 citations), Hedetniemi, S.T. (3,822 citations), and Haynes, T.W. (2,896 citations) highlight their foundational contributions to domination set characterizations and combinatorial optimization \cite{4, 5}. Apparently, most of the researchers dealing with domination in graphs often cited papers that involved foundation concepts such as domination number and bounds, minimal dominating sets, and foundational variations of domination in graphs, among others \cite{5, 31}. These findings suggest that while certain researchers lead in publication count, citation analysis identifies influential works that have shaped the field over time, emphasizing the long-term impact of specific studies rather than sheer volume.
\vspace{5pt}
\par 
The co-authorship network analysis reveals a high level of collaboration among leading researchers, with Sheikhholeslami, S.M. (TLS = 562), Henning, M.A. (TLS = 508), and Chellali, M. (TLS = 296) demonstrating the strongest research connectivity. The presence of well-connected clusters suggests that domination in graphs is largely developed through collaborative efforts rather than isolated contributions. In fact, a collaboration of researchers with different specializations is vital in exploring different aspects of the field of domination in graphs to tackle complex ideas and generate rigorous network structures. At the country level, the United States (TLS = 894) exhibits the strongest research collaborations, followed by Iran (TLS = 558), South Africa (TLS = 503), and China (TLS = 492). This indicates that domination theory research is driven by both Western and emerging research hubs, with increasing contributions from Asian and Middle Eastern countries.
\vspace{5pt}
\par
The co-occurrence network of keywords provides insights into dominant research themes and their interconnections. The first cluster (red) emphasizes computational and algorithmic approaches, suggesting a strong focus on graph algorithms, complexity analysis, and optimization techniques \cite{16, 17}. The second cluster (blue) reflects theoretical investigations into domination numbers and graph invariants, reinforcing that foundational graph properties remain a core research area \cite{18}. The third cluster (green) highlights domination variants, such as total domination and secure domination, which are increasingly applied to network security and real-world system modeling \cite{18, 19}. Finally, the fourth cluster (yellow) suggests a growing interest in binary operations in graphs, indicating a research shift towards how domination properties evolve under structural transformations. This shift aligns with emerging applications in large-scale networks and distributed computing \cite{20}. 
\vspace{5pt}
\par
The citation network analysis confirms that certain foundational studies continue to shape domination research, with Guha (1998) (940 citations), Lund (1994) (613 citations), and Stojmenovic (2002) (605 citations) remaining highly referenced in recent works. The density visualization highlights these papers as core references in the field, while the overlay visualization reveals a shift toward recent studies on binary graph operations, domination in temporal graphs, and deep-learning domination algorithms \cite{21}. The presence of yellow-green nodes in the overlay visualization suggests a transition from traditional theoretical approaches to more applied and computationally intensive studies, reflecting the increasing integration of domination theory with emerging technologies \cite{22, 23}.
\vspace{5pt}
\par 
The co-citation network analysis reveals strong interconnections between foundational studies, demonstrating the evolution of research in domination in graph theory from theoretical foundations to computational applications. The high co-citation strength among domination-related works underscores the enduring significance of classical domination parameters and their extensions in modern studies \cite{24}. The presence of co-cited references in domination variants and computational complexity suggests an increasing integration of domination theory with algorithmic approaches and optimization techniques, particularly in network security, bioinformatics, and combinatorial optimization \cite{25, 26, 27}. The emergence of seven distinct research clusters highlights the diversification of domination studies, ranging from purely theoretical investigations to algorithmic and applied methodologies. The strong co-citation links to complexity-related works indicate that domination problems are frequently studied in relation to NP-hardness and approximation algorithms, reinforcing their role in computational complexity and algorithm design \cite{28, 29, 36, 37}. The integration of domination theory with optimization and network science suggests its expanding applications beyond traditional graph theory, particularly in communication networks, social network analysis, and large-scale computational models \cite{20, 30}. These findings indicate a shifting research focus toward computational and data-driven methodologies, paving the way for future studies that incorporate machine learning, artificial intelligence, and large-scale network modeling to further advance domination-based graph research.
\section{Conclusion}
The evolution of domination in graph theory from a purely theoretical construct to a computationally intensive research domain emphasizes its growing significance across various disciplines. This study highlights the increasing relevance of domination theory, particularly in network security, wireless communication, and combinatorial optimization, as evidenced by the surge in research output after 2010. The influence of prolific researchers and foundational works continues to shape theoretical advancements, while strong international collaborations drive the field’s development. The findings indicate a shift toward data-driven methodologies, with emerging studies exploring deep-learning algorithms and large-scale graph transformations. To advance this field, future research should focus on expanding domination-based models to accommodate evolving network structures, particularly in artificial intelligence, cybersecurity, and large-scale computational systems. The integration of machine learning and deep-learning techniques can enhance graph optimization and real-time network analysis, while further exploration of domination in temporal and dynamic graphs will address challenges in social networks, biological systems, and infrastructure modeling. Strengthening international research collaborations will facilitate knowledge exchange and interdisciplinary innovations. Additionally, periodic bibliometric analyses are essential to track research progress and ensure that theoretical developments align with modern computational applications and real-world problem-solving needs.

\bibliographystyle{amsplain}
\bibliography{biblio}

\end{document}